\documentclass[hyper]{JHEP} 
\newcommand\fverb{\setbox\pippobox=\hbox\bgroup\verb}
\newcommand\fverbdo{\egroup\medskip\noindent%
            \fbox{\unhbox\pippobox}\ }

\newcommand\fverbit{\egroup\item[\fbox{\unhbox\pippobox}]}
\newbox\pippobox
\title{Semiclassical Strings in Supergravity PFT}
\author{Aritra Banerjee, Sagar Biswas and Kamal L. Panigrahi\\
Department of Physics, \\
Indian Institute of Technology Kharagpur,\\
Kharagpur-721 302, INDIA \\
Email: \email{aritra,sbiswas,panigrahi@phy.iitkgp.ernet.in}}
\vskip .2in \abstract{Puff Field Theory (PFT) is an example of a
non-local field theory which arises from a novel embedding of
D-branes in Melvin universe. We study several rotating and
pulsating string solutions of the F-string equations of motion in
the supergravity dual of the PFT. Further, we find a PP-wave
geometry from this nonlocal spacetime by applying a Penrose limit
and comment on its similarity with the maximally supersymmetric
PP-wave background.} 

\begin{document}
\section{Introduction and Summary}
It is not uncommon to find examples of Quantum Field Theories
(QFT) which violate Lorentz invariance in high energy limit. These
theories might play a crucial role in understanding physics beyond
the Standard Model of partical physics. In the context of string theory, for example,
a few Lorentz violating theories are constructed from the local
deformation of the $N=4$ Super Yang-Mills(SYM) theory.
The UV-completeness of such theories are recovered by constraining
the conformal dimensions of such deformation operators, although
at IR limit, action for these theories can approach to that of
$N=4$ SYM theory. Example of such a theory includes $N=4,\>\>
U(N)$ SYM on a space of noncommutative $\mathcal{R}^4$ \cite{Connes:1997cr}, which in
the IR limit looks like N=4 SYM deformed by an operator of conformal
dimension $\Delta=6$ , breaking the Lorentz group $SO(3,1)$ to
$SO(2)\times SO(1,1)$. The non-commutativity introduces a
fundamental linear non-locality into the construction of such a
theory. It is worth mentioning that in many of these theories the
fundamental particles can become extended non-local objects,
making them intriguing for string theorists. It is therefore,
interesting to explore such possible extensions of field theories
that incorporate the violation of Lorentz invariance at some
typical mass scales.

Puff Field Theory (PFT) \cite{Ganor:2006ub} is such an example of a Lorentz violating
non-local field theory. The idea follows the construction of
Non-Commutative SYM (NCSYM) by Douglas and Hull
\cite{Douglas:1997fm}. In NCSYM we consider $n$-coincident $D0$
branes in type IIA string theory compactified on a small $T^2$.
This theory  is T-dual to type IIA on a large $T^2$ with $n$ D2
branes. But this T-duality does not simply map the small $T^2$ to
a large one if a NS NS 2-form flux B$_{\mu\nu}$ is turned on along
$T^2$ as an obstruction. It was argued by Douglas and Hull that
the D2-branes in this setting will be described by non local
interactions in the NCSYM. The construction of PFT is a variant of
such a small/large volume duality. Now consider a Kaluza-Klein
particle with $n$ units of momentum in type IIA string theory
compactified on a $T^3$. An appropriate U-duality transformation
transforms this setting into $n$ D3-branes on type IIB theory
compactified on large $T^3$. Instead of B$_{\mu\nu}$ flux as in
the previous case, we give a geometrical twist that will prevent
U-duality from producing  type IIB on a large $T^3$. It has been
argued in \cite{Ganor:2006ub} that in the low energy limit the
Kaluza-Klein particle is described by a decoupled non-local field
theory that breaks Lorentz symmetry $SO(3,1)$ but preserves
rotational invariant group in three dimensions, $SO(3)$. This
conjectured field theory, where the particle carrying a R-charge
now expands to occupy a D3 brane worldvolume proportional to the
R-charge and the dimensionful deformation parameter, is termed as
Puff Field Theory (PFT). Nothing is known about the explicit
lagrangian form of PFT, but the supergravity description of PFT
can be obtained from the non-trivial embedding of D-brane geometry
in a Melvin universe, as done in \cite{Ganor:2007qh}. The result
is a type IIB supergravity background supported by a 4-form RR
flux and a constant dilaton. While constructing the supergravity
dual background of PFT it has been demanded that the setting
should preserve a few of the supersymmetries to avoid instability
altogether. It has also been argued that the supersymmetry
preservation for this field theory will depend on the nature of
symmetry of the deformation parameter. This can in turn be fixed
by choosing the geometrical twist accordingly.

Now we can see that the background dual to PFT looks incredibly
complex. But, in this work we find that the near horizon
geometry of the background, under Penrose limit, reduces to
the PP-wave of $AdS_5 \times S^5$. This result prompts us to look
for solutions of the F-string equations of motion in this
background in the semiclassical limit. In the context of AdS/CFT
duality, string solutions in the semiclassical limit have proved to
be of key importance in exploring various aspects of the
correspondence. According to AdS/CFT correspondence
\cite{Maldacena:1997re}, \cite{Gubser:1998bc}, \cite{Witten:1998qj}
quantum closed string states in bulk should be dual to local
operators on the boundary. This state-operator matching can be
tractable only in large angular momentum limit, on both sides of
the duality \cite{Berenstein:2002jq}, \cite{Gubser:2002tv},
\cite{Beisert:2005tm}, \cite{Minahan:2002ve},
\cite{Beisert:2003xu}, \cite{Beisert:2003yb},
\cite{Kazakov:2004qf}, \cite{Beisert:2004hm},
\cite{Arutyunov:2004vx}, as both the string theory and the gauge
theory are integrable in the semiclassical limit, see for example
\cite{Pohlmeyer:1975nb}, \cite{Minahan:2002rc},
\cite{Tseytlin:2004xa}, \cite{Hayashi:2007bq},
\cite{Okamura:2008jm}.  In this connection a large number of
rotating and pulsating string solutions have been studied in
various string theory backgrounds, see for example,
\cite{Kruczenski:2004wg}, \cite{Bobev:2005cz},
\cite{Ryang:2004tq}, \cite{Dimov:2004xi}, \cite{Smedback:1998yn},
\cite{Hofman:2006xt}, \cite{Dorey:2006dq}, \cite{Chen:2006gea},
\cite{Bobev:2006fg}, \cite{Kruczenski:2006pk}, \cite{Chen:2006gq},
\cite{Hirano:2006ti}, \cite{Ryang:2006yq},
\cite{Maldacena:2006rv}, \cite{Kluson:2007qu},
\cite{Ishizeki:2007we}, \cite{Bobev:2007bm}, \cite{Hofman:2007xp},  \cite{Ishizeki:2007kh}, \cite{Dorey:2007an},
\cite{Kruczenski:2008bs}, \cite{Lee:2008sk}, \cite{David:2008yk},
\cite{Kluson:2008gf}, \cite{Lee:2008ui}, \cite{Biswas:2011wu},
\cite{Panigrahi:2011be}, \cite{Biswas:2012wu},
\cite{Biswas:2013ela}. Here, we try to extract some simple
solutions following results from these works.

In the case of our background, we expand it in the near horizon
limit keeping only $AdS_5 \times S^5$ plus the leading order
deformation terms, containing the mixing of coordinates from both
$AdS$ and sphere part. It it already shown in \cite{Ganor:2006ub} that this leading
order term, in the dual gauge theory corresponds to a deformation
operator of conformal dimension $\Delta=7$ to N=4 SYM. That is, in
the low energy limit the total Lagrangian can be written as

\begin{equation}
 L=L_{N=4}+\eta {\mathbf{O}}^{(7)}+...
\end{equation}
Where $\eta$ is the dimensionful deformation parameter. Thus, we
choose to ignore the higher order deformation terms in our metric
and study a general class of rotating string solutions in some
approximation. We find that the dispersion relation among various
conserved quantities differ slightly from that of general $AdS_n
\times S^n$. Next we study a class of
solutions both rotating and pulsating in this background. Such
kind of string states are expected to be dual to highly excited
sigma model operators. As oscillation number is a quantum
adiabatic invariant, the series relation of energy in terms of
oscillation number and other conserved quantities is presented as
the solution to characterize the dynamics of these string states.

The rest of the paper is organized as follows. In section-2 we
note down the supergravity description of PFT and take the
appropriate near horizon limit for studying the rotating string
solutions. In section-3, we study penrose limit of the
supergravity dual background of PFT. Section-4 is devoted to the
study of rigidly rotating strings in this background. We present
the regularized dispersion relations among various conserved
charges corresponding to the string motion. We also present
solutions for strings which are both rotating and pulsating in the
above background. Finally, in section-5 we conclude with some
comments.

\section{Supergravity description of PFT} Following \cite{Ganor:2007qh} we
know the supergravity dual background of PFT is given by the
following metric and the 4-form field as,
\begin{eqnarray}
    \frac{ds^2}{\alpha^{\prime}} &=& K^{\frac{1}{2}} \Big(-H^{-1}dt^2 + dU^2 + U^2ds_2^2
    + \sum_{i=8}^9 dY_i^2  \Big) \nonumber \\ &+& K^{-\frac{1}{2}} \Big( \sum_{i=1}^3dx_i^2 + HU^2(d\phi
    + \mathcal{A} + \Delta^3H^{-1}dt)^2\Big) \ , \nonumber \\ \frac{A}{{\alpha^{\prime}}^2}
    &=& K^{-1}(-dt+U^2\Delta^3(d\phi+\mathcal{A}))\wedge dx_1\wedge dx_2 \wedge dx_3 \ , \nonumber \\
    e^{\phi} &=& g_{IIB}=2\pi g^2_{YM} \ ,
\end{eqnarray}
where the harmonic functions $H$ and $K$ are,
\begin{eqnarray}
    H  &= & \frac{4\pi g_{IIB}N}{(U^2+||Y||^2)^2} \ , \>\>\>
    K =  \frac{4\pi g_{IIB}N}{(U^2+||Y||^2)^2} + \Delta^6U^2 \ ,
\end{eqnarray}
also $ds^2_2=\frac{1}{4}(d\theta^2 + \sin^2\theta d\varphi^2)$ is
the ``Fubini- Study" metric and
$\mathcal{A}=-\frac{1}{2}(1-\cos\theta)d\varphi$ is the connection
of Hopf fibration. Note that to obtain this background one needs
to take the decoupling limit $\alpha^{\prime} \rightarrow 0$.
However, in this limit the value of $\Delta^3=\eta
{\alpha^{\prime}}^2$ is held fixed for large value of deformation
parameter $\eta$.

Now, considering $U=V\cos\zeta$ and $||Y||=V\sin\zeta$ i.e. $Y_8=V\sin\zeta\cos\psi$
and $Y_9=V\sin\zeta\sin\psi$  we can rewrite the metric and 4-form as follows \cite{Ganor:2007qh},
\begin{eqnarray}
    \frac{ds^2}{\alpha^{\prime}} &=& K^{\frac{1}{2}}(-K^{-1}dt^2 + dV^2 + V^2d\zeta^2
    + V^2 \sin^2\zeta d\psi^2) + \frac{1}{4}K^{\frac{1}{2}}V^2 \cos^2\zeta d\theta^2 \nonumber \\
    &+& \frac{1}{4}K^{-\frac{1}{2}}V^2 \cos^2\zeta (K\sin^2\theta + H(1-\cos\theta)^2)d\varphi^2
    + K^{-\frac{1}{2}}HV^2\cos^2\zeta d\phi^2  \nonumber \\ &+&  K^{-\frac{1}{2}}
    \sum_{i=1}^3dx_i^2 + 2K^{-\frac{1}{2}}V^2\cos^2\zeta \Delta^3 dt d\phi  \nonumber \\
    &-& K^{-\frac{1}{2}}HV^2\cos^2\zeta (1-\cos\theta)d\phi d\varphi -
    K^{-\frac{1}{2}}V^2\cos^2\zeta \Delta^3(1-\cos\theta)dt d\varphi \ , \nonumber \\
    \frac{A}{{\alpha^{\prime}}^2} &=& K^{-1}(-dt + \Delta^3V^2\cos^2\zeta(d\phi-\frac{1}{2}(1-\cos\theta)
    d\varphi))\wedge dx_1 \wedge dx_2 \wedge dx_3 \ , \nonumber \\ e^{\phi} &=& 2\pi g^2_{YM} \ ,
\end{eqnarray}
with  $K=H+\Delta^6V^2\cos^2\zeta$, $H=\frac{8\pi^2g^2_{YM}N}{V^4}$. Now we want to take near horizon limit on this full generalised
metric. Note that in near horizon limit (i.e. $V \rightarrow 0$),
$H=\frac{C^2}{V^4} \approx K$, where
$C^2=8\pi^2g^2_{YM}N$, and we have kept terms upto $V^4$. The resulting
metric and the four form field is,
\begin{eqnarray}
    \frac{ds^2}{\alpha^{\prime}} &=& \frac{V^2}{C}(-dt^2 + \sum_{i=1}^3dx_i^2)
 + C \frac{dV^2}{V^2} + \frac{2\Delta^3 V^4}{C}\cos^2\zeta dt(d\phi
- \sin^2(\frac{\theta}{2})d\varphi) \nonumber \\ &+& C[d\zeta^2 +
\sin^2\zeta d\psi^2 + \cos^2\zeta \{(\frac{d\theta}{2})^2 +
d\phi^2 + \sin^2(\frac{\theta}{2})d\varphi^2 -
2\sin^2(\frac{\theta}{2})d\phi d\varphi\}] \ , \nonumber \\
\frac{A}{{\alpha^{\prime}}^2} &=& -\frac{V^4}{C^2}dt\wedge dx_1
\wedge dx_2 \wedge dx_3 \ .
\end{eqnarray}
Now making the following change of variables,
\begin{eqnarray}
    \theta &=& 2\theta, \>\>\> \varphi = \phi_1 - \phi_2, \>\>\> \phi = \phi_1,
    \>\>\> \zeta = \zeta - \frac{\pi}{2} \nonumber \ ,
\end{eqnarray}
 we get,
\begin{eqnarray}
    \frac{ds^2}{\alpha^{\prime}} &=& \frac{V^2}{C}(-dt^2 + \sum_{i=1}^3dx_i^2)
+ C \frac{dV^2}{V^2} + \frac{2\Delta^3 V^4}{C}\sin^2\zeta
dt(\cos^2\theta d\phi_1 + \sin^2\theta d\phi_2) \nonumber \\ &+&
C[d\zeta^2 + \cos^2\zeta d\psi^2 + \sin^2\zeta(d\theta^2 +
\cos^2\theta d\phi_1^2 + \sin^2\theta d\phi_2^2)] \ , \nonumber \\
\frac{A}{{\alpha^{\prime}}^2} &=& -\frac{V^4}{C^2}dt\wedge dx_1
\wedge dx_2 \wedge dx_3 \ . \label{metric 1}
\end{eqnarray}
This is the metric we are interested in taking a Penrose limit.

\section{Penrose limit}
In this section we would like to find out a PP-wave metric by applying
Penrose limit on the background (\ref{metric 1}).
To take Penose limit on (\ref{metric 1}), we start with a
null geodesic in  ($t$, $V$ ,$\psi$) plane following \cite{Blau:2002dy}.
Keeping other coordinates fixed, the metric becomes
\begin{equation}
    \frac{ds^2}{\alpha^{\prime}} = C[-V^2dt^2 + \frac{dV^2}{V^2} + d\psi^2] \ . \label{geodesic}
\end{equation}
To change the coordinates from $(t,V,\psi)$ to $(u,v,y)$, which are more
suitable to adapt the null geodesic,  we use the following transformation
\begin{eqnarray}
    dV &=& \sqrt{1-l^2V^2}du \ , \nonumber \\ dt &=& \frac{du}{V^2} + ldy - dv \ ,
    \nonumber \\ d\psi &=& ldu + dy \ , \label{trans}
\end{eqnarray}
where $l=\frac{J}{E}$, $J$ and $E$ respectively are angular momentum and energy along
the geodesic (\ref{geodesic}). Substituting (\ref{trans}) in (\ref{metric 1}), and making the change of coordinates,
\begin{eqnarray}
    u &=& u, \>\>\> v=\frac{v}{C}, \>\>\> y=\frac{y}{\sqrt{C}}, \>\>\> x_i=\frac{x_i}{\sqrt{C}},
    \>\>\> \zeta=\frac{z}{\sqrt{C}}, \>\>\> \Omega_3=\Omega_3, \nonumber
\end{eqnarray}
followed by a large $C$ limit, the metric and the field strength reduces to,
\begin{eqnarray}
    \frac{ds^2}{\alpha^{\prime}} &=& 2dudv - z^2l^2du^2 + (1-l^2V^2)dy^2
+ V^2\sum_{i=1}^3dx_i^2 + dz^2 + z^2d\Omega_3^2 \ , \nonumber \\ F &=&
dA = -4V^3l\sqrt{1-l^2V^2}du\wedge dy\wedge dx_1\wedge dx_2\wedge
dx_3 \ .
\end{eqnarray}
Again rescaling $u \rightarrow \mu u$ and $v \rightarrow \frac{v}{\mu}$, we get,
\begin{eqnarray}
    \frac{ds^2}{\alpha^{\prime}} &=& 2dudv - \mu^2z^2l^2du^2 + (1-l^2V^2)dy^2
+ V^2\sum_{i=1}^3dx_i^2 + d\bar{z}^2 \ , \nonumber \\ F_{uyx_1x_2x_3}
&=& -4\mu V^3l\sqrt{1-l^2V^2}.
\end{eqnarray}
where $d\bar{z}^2=dz^2 + z^2d\Omega_3^2$. This is the Rosen form of the PP wave.
To convert this into Brinkman form we make the following substitution,
\begin{eqnarray}
    u &=& u, \>\>\> y=\frac{y}{\sqrt{1-l^2V^2}}, \>\>\> x_i= \frac{x_i}{V},
\>\>\> \bar{z} = \bar{z}, \nonumber \\ v &=& v +
\frac{1}{4}\Big[\frac{\partial_u(1-l^2V^2)}{1-l^2V^2}y^2 +
\frac{\partial_u(V^2)}{V^2}\sum_{i=1}^3x_i^2\Big]  \ ,
\end{eqnarray}
Substituting these we get the Brinkman form of the PP-wave as,
\begin{eqnarray}
    \frac{ds^2}{\alpha^{\prime}} &=& 2dudv + (F_1y^2 + F_2x_i^2 - \mu^2z^2l^2)du^2
+ dy^2 + \sum_{i=1}^3dx_i^2 + d\bar{z}^2 \ , \nonumber \\
F_{uyx_1x_2x_3} &=& -4\mu l,
\end{eqnarray}
where \begin{eqnarray} F_1=\frac{1}{2}[\partial_u
\{\frac{\partial_u(1-l^2V^2)}{1-l^2V^2}\} +
\frac{1}{2}\{\frac{\partial_u(1-l^2V^2)}{1-l^2V^2}\}^2] \ ,
\nonumber
\\
F_2=\frac{1}{2}[\partial_u \{\frac{\partial_u(V^2)}{V^2}\} +
\frac{1}{2}\{\frac{\partial_u(V^2)}{V^2}\}^2].
\end{eqnarray}
This form is similar to the form that is obtained by taking a
Penrose limit on the geometry of a stack of N D3-branes in the near horizon
limit. String propagation in this background has been studied in detail
\cite{Berenstein:2002jq}. The main output of this section is
that the very complicated metric (\ref{metric 1}) reduces to a well known
form in Penrose limit. That signifies
that when we consider the deformation term to be small, the local
geometry will behave like
$AdS_5\times S^5$ to a local observer on the geodesic mentioned in this section.
In the next section we will be interested in finding solutions of string equation of motion in
semiclassical limit in the background (\ref{metric 1}).
\section{Semiclassical String solutions}
If we neglect $V^4$ term in (\ref{metric 1}), then the metric simply
takes the form of $AdS_5\times S^5$, for which the rigidly
rotating string solutions are well studied. It would be
interesting if we can find string solutions by keeping the first order term in $V^4$.
By rescaling, $t \rightarrow \Delta C^{\frac{2}{3}}t$, $x_i
\rightarrow \Delta C^{\frac{2}{3}}x_i$ and substituting
$V=\frac{1}{\Delta}WC^{\frac{1}{3}}$, we get,
\begin{eqnarray}
    \frac{ds^2}{\alpha^{\prime}} &=& C\Big[W^2(-dt^2 + \sum_{i=1}^3dx_i^2) +  \frac{dW^2}{W^2}
    + 2W^4\sin^2\zeta dt(\cos^2\theta d\phi_1 + \sin^2\theta d\phi_2) \nonumber \\
    &+& d\zeta^2 + \cos^2\zeta d\psi^2 + \sin^2\zeta(d\theta^2 + \cos^2\theta d\phi_1^2
    + \sin^2\theta d\phi_2^2)\Big] \ , \nonumber \\
    \frac{A}{{\alpha^{\prime}}^2} &=& -C^2W^4dt\wedge dx_1 \wedge dx_2 \wedge dx_3 \ . \label{metric 2}
\end{eqnarray}
It is very hard to solve the  equations of motion for the fundamental string
in the above background (\ref{metric 2}), since they lead to highly non-linear
coupled differential equations. However, we
can simplify and consider a less general geometry than (\ref{metric 2})
 by putting $W=W_0$ and $\theta=\theta_0$. For these values, the metric (\ref{metric 2}) becomes,
\begin{eqnarray}
    \frac{ds^2}{\alpha^{\prime}} &=& C\Big[W^2_0(-dt^2 + \sum_{i=1}^3dx_i^2)
    + 2W_0^4\sin^2\zeta dt(\cos^2\theta_0 d\phi_1 + \sin^2\theta_0 d\phi_2) \nonumber \\
    &+& d\zeta^2 + \cos^2\zeta d\psi^2 + \sin^2\zeta(\cos^2\theta_0^2d\phi_1^2
    + \sin^2\theta_0 d\phi_2^2)\Big] \ , \label{metric 3}
\end{eqnarray}
where $W_0$ and $\theta_0$ are constants. In the following
analysis we will keep the terms upto $\mathcal{O}({W_0}^4)$ only.
It can be noted that making the coordinates
$W$ and $\theta$ constant will certainly impose some non-trivial
constraints on the string solutions in this background. We will however,
show that these constraints merely reduce to some relations between
the various constants mentioned in the wordsheet embedding of our choice.
\subsection{Rigidly Rotating Strings}
 We start our analysis by writing down the Polyakov action of the
 F-string in the background (\ref{metric 3}),
\begin{equation}
S=-\frac{1}{4\pi\alpha^{\prime}}\int d\sigma d\tau
[\sqrt{-\gamma}\gamma^{\alpha \beta}g_{MN}\partial_{\alpha} X^M
\partial_{\beta}X^N ]\ ,
\end{equation}
where
$\gamma^{\alpha \beta}$ is the world-sheet metric . Under conformal gauge (i.e.
$\sqrt{-\gamma}\gamma^{\alpha \beta}=\eta^{\alpha \beta}$) with
$\eta^{\tau \tau}=-1$, $\eta^{\sigma \sigma}=1$ and $\eta^{\tau
\sigma}=\eta^{\sigma \tau}=0$, the Polyakov action in the above
background takes the form,
\begin{eqnarray}
S &=& -\frac{\sqrt{\lambda}}{4\pi}\int d\sigma
d\tau\Big[W_0^2\{-({t^{\prime}}^2-\dot{t}^2) +
{x_i^{\prime}}^2-\dot{x_i}^2\} + {\zeta^{\prime}}^2-\dot{\zeta}^2
\nonumber \\ &+& \cos^2\zeta({\psi^{\prime}}^2-\dot{\psi}^2) +
\sin^2\zeta\{{\cos^2\theta_0({\phi_1^{\prime}}^2-\dot{\phi_1}^2)+
\sin^2\theta_0(\phi_2^{\prime}}^2-\dot{\phi_2}^2)\} \nonumber \\
&+& 2W_0^4\sin^2\zeta\{\cos^2\theta_0(t^{\prime}\phi_1^{\prime} -
\dot{t}\dot{\phi_1}) + \sin^2\theta_0(t^{\prime}\phi_2^{\prime} -
\dot{t}\dot{\phi_2}) \} \Big] \ ,
\end{eqnarray}
where `dots' and `primes' denote the derivative with respect to
$\tau$ and $\sigma$ respectively, also 't Hooft coupling $\sqrt{\lambda}=C$.
For studying the rigidly rotating strings in this background we choose the following  ansatz,
\begin{eqnarray}
t &=& \tau + h_0(y) ,\>\> x_i=\nu_i(\tau+h_i(y)), \> i=1,2,3, \>\>\>
\zeta=\zeta(y),   \nonumber \\ \phi_1 &=& \omega_1(\tau+g_1(y)), \>\>\>
\phi_2=\omega_2(\tau+g_2(y)), \>\>\> \psi=\omega_3(\tau+g_3(y))
\ , \label{ansatz}
\end{eqnarray}
where $y=\sigma-v\tau$. Variation of the action with respect to
$X^M$ gives us the following equation of motion
\begin{eqnarray}
2\partial_{\alpha}(\eta^{\alpha \beta} \partial_{\beta}X^Ng_{KN})
&-& \eta^{\alpha \beta} \partial_{\alpha} X^M \partial_{\beta}
X^N\partial_K g_{MN} =0 \ ,
\end{eqnarray}
and variation with respect to the metric gives the two Virasoro
constraints,
\begin{eqnarray}
g_{MN}(\partial_{\tau}X^M \partial_{\tau}X^N +
\partial_{\sigma}X^M \partial_{\sigma}X^N)&=&0 \ , \nonumber \\
g_{MN}(\partial_{\tau}X^M \partial_{\sigma}X^N)&=&0 \ .
\end{eqnarray}
Next we have to solve these equations by the ansatz we have
proposed above in eqn. (\ref{ansatz}). Solving for $t$, $\phi_1$ and $\phi_2$ we
get,
\begin{eqnarray}
    -\frac{\partial h_0}{\partial y} + \omega_1W_0^2\cos^2\theta_0\sin^2\zeta
\frac{\partial g_1}{\partial y} &+&
\omega_2W_0^2\sin^2\theta_0\sin^2\zeta \frac{\partial
g_2}{\partial y} \nonumber \\ &=&
\frac{1}{1-v^2}[c_4-vW_0^2\sin^2\zeta\{\omega_1\cos^2\theta_0 +
\omega_2\sin^2\theta_0\}] \ , \nonumber \\ W_0^4\sin^2\zeta
\frac{\partial h_0}{\partial y} + \omega_1\sin^2\zeta
\frac{\partial g_1}{\partial y} &=& \frac{1}{1-v^2}[c_5 -
v\sin^2\zeta(\omega_1 + W_0^4)] \ , \nonumber \\ W_0^4\sin^2\zeta
\frac{\partial h_0}{\partial y} + \omega_2\sin^2\zeta
\frac{\partial g_2}{\partial y} &=& \frac{1}{1-v^2}[c_6 -
v\sin^2\zeta(\omega_2 + W_0^4)] \ , \label{eqn}
\end{eqnarray}
where $c_4$, $c_5$ and $c_6$ are integration constants. Solving (\ref{eqn}),
 we get
\begin{eqnarray}
    \frac{\partial h_0}{\partial y} &=& \frac{1}{1-v^2}[W_0^2(c_5\cos^2\theta_0 + c_6\sin^2\theta_0)-c_4] \ , \nonumber \\
\frac{\partial g_1}{\partial y} &=&
\frac{1}{1-v^2}\Big[\frac{1}{\omega_1}\{\frac{c_5}{\sin^2\zeta} - W_0^4(v-c_4)\}-v\Big]
\ , \nonumber \\ \frac{\partial g_2}{\partial y} &=&
\frac{1}{1-v^2}\Big[\frac{1}{\omega_2}\{\frac{c_6}{\sin^2\zeta} - W_0^4(v-c_4)\}-v\Big] \ .
\end{eqnarray}
Solving for $\psi$ and $x_i$, respectively we get,
\begin{eqnarray}
    \frac{\partial g_3}{\partial y} &=& \frac{1}{1-v^2}[\frac{c_7}{\cos^2\zeta}-v], \>\>\>
    \frac{\partial h_i}{\partial y} = c_i \ ,
\end{eqnarray}
where $c_7$ and $c_i$, $(i=1,2,3)$ are integration constants. As
discussed before, putting $W$ and $\theta$ as constants generates
some confining constraint equations from the equations of motion
for $W$ and $\theta$. These constraint equations in this case can
be written as,
\begin{eqnarray}
    4W_0^2\sin^2\zeta(\omega_1\cos^2\theta_0 + \omega_2\sin^2\theta_0)
    &=& (W_0^2d_1 + v - c_4)(3W_0^2d_1-v+c_4)  \nonumber \\
    &+& 1 - (1-v^2)\nu_i^2\{(1-vc_i)^2 - c_i^2\} \nonumber \\
    \frac{c_6^2-c_5^2}{\sin^4\zeta} + \omega_1^2 - \omega_2^2 &=& 2W_0^4(\omega_2 - \omega_1) \ , \label{constraint 1}
\end{eqnarray}
where $d_1=c_5\cos^2\theta_0 + c_6\sin^2\theta_0$. These
constraints (\ref{constraint 1}) will imply $\zeta=$ constant,
which is a trivial solution. To have non-trivial solution for
strings in this supergravity PFT background, we must put
\begin{eqnarray}
    \omega_1\cos^2\theta_0 + \omega_2\sin^2\theta_0 &=& 0,\>\>\> c_5 = c_6 \ .
    \label{constraint 2}
\end{eqnarray}
Using (\ref{constraint 2}), (\ref{constraint 1}) can be put in the form,
\begin{eqnarray}
     (W_0^2c_5 + v - c_4)(3W_0^2c_5-v+c_4) + 1 &=&  (1-v^2)\nu_i^2\{(1-vc_i)^2 - c_i^2\}
     \nonumber \\
     \omega_1 + \omega_2 &=& -2W_0^4 \ . \label{constraint 3}
\end{eqnarray}
Since these above equations confine our parameter space
non-trivially, we have to be careful in our approach for analyzing
string solutions. As a check we can see that using the conditions
mentioned in (\ref{constraint 2}) and solving for $\zeta$ we get,
\begin{equation}
   (1-v^2)^2\frac{\partial^2\zeta}{\partial y^2} = \sin\zeta \cos\zeta [
   \frac{c_5^2}{\sin^4\zeta} -\frac{\omega_3^2c_7^2}{\cos^4\zeta} - \omega^2] \ , \label{zeta 1}
\end{equation}
where $\omega^2=\omega_1^2\cos^2\theta_0 +
\omega_2^2\sin^2\theta_0 - \omega_3^2$. Integrating
(\ref{zeta 1}), we get,
\begin{equation}
    (1-v^2)^2\Big(\frac{\partial \zeta}{\partial y}\Big)^2= -\frac{c_5^2}{\sin^2\zeta} -
    \frac{\omega_3^2c_7^2}{\cos^2\zeta} - \omega^2\sin^2\zeta + c_8 , \label{zeta 2}
\end{equation}
where $c_8$ is integration constant. For self consistency of the
solution, the equation (\ref{zeta 2}) will have to be properly
supplemented by the two Virasoro constraints.

The Virasoro constraint $g_{MN}(\partial_{\tau}X^M \partial_{\sigma}X^N)=0$ in this case
will become,
\begin{eqnarray}
    (1 &-& v^2)^2 \Big(\frac{\partial \zeta}{\partial y}\Big)^2 = W_0^2(W_0^2c_5 - c_4)^2 -
    \frac{1-v^2}{v}W_0^2(W_0^2c_5 - c_4) + (1-v^2)^2W_0^2\nu_i^2\frac{c_i}{v} \nonumber \\
    &-& (1-v^2)^2W_0^2\nu_i^2c_i^2  - \omega_3^2\cos^2\zeta  - \sin^2\zeta(\omega_1^2\cos^2 \theta_0 +
    \omega_2^2\sin^2\theta_0) \nonumber \\  &-& \frac{c_5^2}{\sin^2\zeta} - \frac{\omega_3^2c_7^2}{\cos^2\zeta}
    + 2W_0^4(v-c_4)c_5 + \frac{1-v^2}{v}W_0^4c_5 + 2W_0^4c_4c_5 + \Big(\frac{1+v^2}{v}\Big)c_7\omega_3^2 \ . \nonumber \\  \label{Virasoro 1}
\end{eqnarray}
Again the Virasoro $g_{MN}(\partial_{\tau}X^M \partial_{\tau}X^N +
\partial_{\sigma}X^M \partial_{\sigma}X^N)=0$ becomes,
\begin{eqnarray}
    (1 &-& v^2)^2 \Big(\frac{\partial \zeta}{\partial y}\Big)^2 =
    W_0^2(W_0^2c_5 - c_4)^2 + \frac{1-v^2}{1+v^2}W_0^2\{1-v^2 -2v(W_0^2c_5 - c_4)\} -
    (1-v^2)^2W_0^2\nu_i^2c_i^2  \nonumber \\ &-&  \frac{(1-v^2)^2}{1+v^2}W_0^2\nu_i^2(1-2vc_i)
    - \omega_3^2\cos^2\zeta  - \sin^2\zeta(\omega_1^2\cos^2\theta_0 + \omega_2^2\sin^2\theta_0) \nonumber \\
    &-& \frac{c_5^2}{\sin^2\zeta} - \frac{\omega_3^2c_7^2}{\cos^2\zeta} + 2W_0^4(v-c_4)c_5 +
    \frac{v(1-v^2)}{1+v^2}W_0^4c_5 + W_0^4c_4c_5 + \Big(\frac{4v}{1+v^2}\Big)c_7\omega_3^2  \ . \nonumber \\  \label{Virasoro 2}
\end{eqnarray}
Subtracting these two Virasoro constraints we get another relation between the constants,
\begin{equation}
    c_7\omega_3^2 + W_0^2\nu_i^2\{(1-v^2)c_i + v\} -W_0^2(v-c_4) + W_0^4c_5\frac{v^2(1-v^2) + c_4v(1+v^2)}{(1-v^2)^2} = 0  \ . \label{constraint 4}
\end{equation}
Note that from (\ref{Virasoro 1}) if we identify
\begin{eqnarray}
    c_8 &=& W_0^2(W_0^2c_5-c_4)^2 + 2W_0^4vc_5 + \frac{1-v^2}{v}W_0^2c_4 \nonumber \\ &+& (1-v^2)^2W_0^2\nu_i^2(\frac{c_i}{v}-c_i^2) -
    \omega_3^2 + \frac{1+v^2}{v}\omega_3^2c_7 \ ,
\end{eqnarray}
then (\ref{Virasoro 1}) is consistent with equation of motion
(\ref{zeta 2}). To summarize, the equations (\ref{constraint 3})
and (\ref{constraint 4}) gives the desired constraint equations
for the string solutions in the background (\ref{metric 3}). Since
these constraints are highly non-linear in the parameters, it can
be clearly stated that our rotating string solutions are valid
only in a highly confined parameter space.

Since we are interested in infinite angular momenta solutions
we can consider the limit, $\frac{\partial \zeta}{\partial y}
\rightarrow 0$ as $\zeta \rightarrow \frac{\pi}{2}$, in (\ref{zeta
2}) implies $c_7=0$ and $c_8=c_5^2 + \omega^2$.
 Substituting this in the above equation we get,
\begin{equation}
\frac{\partial \zeta}{\partial y}=\frac{\omega \cot \zeta}{1-v^2}
\sqrt{\sin^2\zeta - \sin^2\zeta_0} \ ,
\end{equation}
where $\sin\zeta_0=\frac{c_5}{\omega}$.
 Looking at the symmetry of the background (\ref{metric 3}), a number of conserved
 charges can be constructed as follows,
\begin{eqnarray}
E &=& -\int\frac{\partial \mathcal{L}}{\partial \dot{t}}d\sigma =
\frac{\sqrt{\lambda}}{2\pi}\frac{W_0^2}{1-v^2}[(1-v^2+vc_4)]\int d\sigma \ , \nonumber \\ P_i
&=& \int\frac{\partial \mathcal{L}}{\partial \dot{x_i}} d\sigma
=\frac{\sqrt{\lambda}}{2\pi}\nu_iW_0^2(1-vc_i)\int d\sigma \ ,
\nonumber \\ J_{\psi} &=& \int\frac{\partial \mathcal{L}}{\partial
\dot{\psi}} d\sigma
=\frac{\sqrt{\lambda}}{2\pi}\frac{\omega_3}{1-v^2}\int \cos^2\zeta
d\sigma \ , \nonumber \\ J_{\phi_1} &=& \int\frac{\partial
\mathcal{L}}{\partial \dot{\phi_1}} d\sigma
=\frac{\sqrt{\lambda}}{2\pi}\frac{\cos^2\theta_0}{1-v^2}\int[(\omega_1
+ W_0^4)\sin^2\zeta-vc_5] d\sigma \ , \nonumber \\
J_{\phi_2} &=& \int\frac{\partial \mathcal{L}}{\partial
\dot{\phi_2}} d\sigma
=\frac{\sqrt{\lambda}}{2\pi}\frac{\sin^2\theta_0}{1-v^2}\int
[(\omega_2 + W_0^4)\sin^2\zeta-vc_5] d\sigma  \ .  \label{charges}
\end{eqnarray}
Also the deficit angles are given by,
\begin{eqnarray}
    \Delta\phi_1 &=& \omega_1\int\frac{\partial g_1}{\partial y}d\sigma =
\frac{1}{1-v^2}\int\Big[\frac{c_5}{\sin^2\zeta} - W_0^4(v-c_4) -
v\omega_1\Big]d\sigma \ , \nonumber \\ \Delta\phi_2 &=&
\omega_2\int\frac{\partial g_2}{\partial y}d\sigma =
\frac{1}{1-v^2}\int\Big[\frac{c_5}{\sin^2\zeta} - W_0^4(v-c_4)
-v\omega_2\Big]d\sigma \ .
\end{eqnarray}
For our convenience, we will use the combined angular momenta and
deficit angles as,
\begin{eqnarray}
    J_{\phi} &=& J_{\phi_1} + J_{\phi_2} = \frac{\sqrt{\lambda}}{2\pi} \frac{1}{1-v^2} \int(W_0^4\sin^2\zeta - vc_5)
    d\sigma \nonumber \\ \Delta\phi &=& \frac{\Delta\phi_1 + \Delta\phi_2}{2} = \frac{1}{1-v^2}\int[\frac{c_5}{\sin^2\zeta}
    - \frac{c_4(\omega_1 + \omega_2)}{2}]d\sigma \ .
\end{eqnarray}
In what follows, we will find relations among various charges in
different limiting cases. Since some of the charges in
\ref{charges} are divergent, we will use a particular type of
regularization technique to extract the relations.
\subsubsection{Case I : Giant Magnon}
For this case, we choose $c_5=\frac{c_4(\omega_1+\omega_2)}{2}$, and the angle deficit becomes,
\begin{equation}
    \Delta\phi=\frac{2c_5}{\omega} \int_{\zeta_0}^{\frac{\pi}{2}}
\frac{\cos\zeta d\zeta}{\sin\zeta \sqrt{\sin^2\zeta -
\sin^2\zeta_0}} = 2\arccos(\sin\zeta_0) \ ,
\end{equation}
which implies $\sin\zeta_0=\cos\Big(\frac{\Delta\phi}{2}\Big)$.
In this condition the expression of energy and linear momenta $P_i$ can be written as,
\begin{eqnarray}
     E &=& \frac{\sqrt{\lambda}}{\pi} \frac{W_0^2}{\omega}[1-v^2+vc_4]
    \int_{\zeta_0}^{\frac{\pi}{2}} \frac{\sin\zeta d\zeta}{\cos\zeta \sqrt{\sin^2\zeta - \sin^2\zeta_0}}, \nonumber \\
    P_i &=& \frac{\sqrt{\lambda}}{\pi} \frac{W_0^2}{\omega}\nu_i(1-v^2)(1-vc_i)
    \int_{\zeta_0}^{\frac{\pi}{2}} \frac{\sin\zeta d\zeta}{\cos\zeta \sqrt{\sin^2\zeta - \sin^2\zeta_0}} \ .
\end{eqnarray}
 It can be seen that these expressions are divergent. But looking at the other charges in this case, we find
\begin{equation}
    J_{\psi} = \frac{\sqrt{\lambda}}{\pi}\frac{\omega_3}{\omega}\cos\zeta_0 \ ,
\end{equation}
is finite, while the combined angular momentum can be written as,
\begin{equation}
    J_{\phi} =  \frac{\sqrt{\lambda}}{\pi} \frac{W_0^4 - vc_5}{\omega} \int_{\zeta_0}^{\frac{\pi}{2}}
\frac{\sin\zeta d\zeta}{\cos\zeta \sqrt{\sin^2\zeta -
\sin^2\zeta_0}} -
\frac{\sqrt{\lambda}}{\pi} \frac{W_0^4}{\omega}
\int_{\zeta_0}^{\frac{\pi}{2}} \frac{\sin\zeta \cos\zeta
d\zeta}{\sqrt{\sin^2\zeta - \sin^2\zeta_0}} \ .
\end{equation}
 It is clear that $J_{\phi}$ also diverges due to the first integral.
 Now we follow the regularization scheme outlined in, for example \cite{Ryang:2006yq} . Let us define a divergent quantity
\begin{equation}
     \tilde{E} = \frac{W_0^4-vc_5}{W_0^2[1-v^2+vc_4 + \nu_i(1-v^2)(1-vc_i)]}(E + \frac{1}{3} \sum P_i)  .
\end{equation}
So that we can write
\begin{equation}
    \tilde{E} - J_{\phi} = \frac{\sqrt{\lambda}}{\pi}\frac{W_0^4}{\omega}
    \cos \zeta_0 ,
\end{equation}
which is a finite quantity. It can be easily shown that the above mentioned conserved charges
obey a dispersion relation among them of the form
\begin{equation}
   \tilde{E} - J_{\phi} = \sqrt{J_\psi^2 + f(\lambda)
   \sin^2\Big(\frac{\Delta\phi}{2}\Big)} \ ,
\end{equation}
where $f(\lambda) = \frac{\lambda}{\pi^2}
\frac{W^8_0-\omega_3^2}{\omega^2}$. The above relation is
analogous to the two spin giant magnon dispersion relation .

\subsubsection{Case II : Single Spike solution}
For this case, choosing $c_5=\frac{c_4(\omega_1+\omega_2)}{2v^2}$, we see that deficit angle
\begin{equation}
  \Delta\phi = \frac{2c_5}{\omega}[(1-v^2)\int_{\frac{\pi}{2}}^{\zeta_0} \frac{\sin\zeta d\zeta}{\cos\zeta
\sqrt{\sin^2\zeta - \sin^2\zeta_0}} + \int_{\frac{\pi}{2}}^{\zeta_0} \frac{\cos\zeta d\zeta}{\sin\zeta
\sqrt{\sin^2\zeta - \sin^2\zeta_0}}]
\end{equation}
diverges due to the first integral. The energy $E$ and linear momenta $P_i$ also diverges as in
the previous case. Here again we will use the divergent combination of the form,
\begin{equation}
  E + \frac{1}{3}\sum P_i = \frac{\sqrt{\lambda}}{\pi} \frac{W_0^2}{\omega}[1-v^2+vc_i +
  \nu_i(1-v^2)(1-vc_i)] \int_{\frac{\pi}{2}}^{\zeta_0} \frac{\sin\zeta d\zeta}{\cos\zeta
\sqrt{\sin^2\zeta - \sin^2\zeta_0}} \ .
\end{equation}
While the other conserved charges are given by
\begin{equation}
    J_{\phi} =  \frac{\sqrt{\lambda}}{\pi} \frac{W_0^4 - vc_5}{\omega} \int^{\zeta_0}_{\frac{\pi}{2}}
\frac{\sin\zeta d\zeta}{\cos\zeta \sqrt{\sin^2\zeta -
\sin^2\zeta_0}} -
\frac{\sqrt{\lambda}}{\pi} \frac{W_0^4}{\omega}
\int^{\zeta_0}_{\frac{\pi}{2}} \frac{\sin\zeta \cos\zeta
d\zeta}{\sqrt{\sin^2\zeta - \sin^2\zeta_0}} \ ,
\end{equation}
which also is diverging due to the first integral and
\begin{equation}
    J_{\psi} = -\frac{\sqrt{\lambda}}{\pi}\frac{\omega_3}{\omega}\cos\zeta_0 \ .
\end{equation}
is finite as before. Now we can regularise the value of $\Delta\phi$ by subtracting out the divergent part,
\begin{eqnarray}
    (\Delta\phi)_{reg} &=& \Delta\phi - \frac{2\pi c_5(1 - v^2)}{\sqrt{\lambda} W_0^2[1-v^2+vc_i +
    \nu_i(1-v^2)(1-vc_i)]} (E + \frac{1}{3}\sum P_i) \nonumber \\ &=& -2\arccos(\sin\zeta_0) \ , \label{nonspike}
\end{eqnarray}
which implies $\sin\zeta_0 = \cos\Big(\frac{(\Delta\phi)_{reg}}{2}\Big)$. Again we write the regularised
value of $J_{\phi}$ as
\begin{eqnarray}
    (J_{\phi})_{reg} &=& J_{\phi} - \frac{W_0^4 - vc_5}{W_0^2[1-v^2+vc_i + \nu_i(1-v^2)(1-vc_i)]} (E + \frac{1}{3}\sum P_i) \ , \nonumber \\
    &=& \frac{\sqrt{\lambda}}{\pi}\frac{W_0^4}{\omega}\cos\zeta_0 \ .
\end{eqnarray}
We can see that the constants of motion satisfy the following dispersion relation
\begin{equation}
    (J_{\phi})_{reg} = \sqrt{J_{\psi}^2 + f(\lambda)\sin^2\frac{(\Delta\phi)_{reg}}{2}} \
    ,
\end{equation}
where $f(\lambda) = \frac{\lambda}{\pi^2}
\frac{W^8_0-\omega_3^2}{\omega^2}$. This looks like the spiky
string dispersion relation presented in \cite{Ishizeki:2007we}.
\subsection{Rotating and Pulsating Strings with two equal spins}
In this section we will focus on a class of `long' semiclassical
strings which are both pulsating and rotating in the background
(\ref{metric 1}). Here we follow the simple procedure as done in,
for example \cite{Park:2005kt} for our analysis.
\footnote{Recently more generalized rotating and pulsating strings
have been studied in \cite{Pradhan:2013sja}} We again put $W=W_{0}$ and
$\theta=\frac{\pi}{4}$ for simplicity in the metric and keep terms upto
$W_0^4$ keeping in with our approximation as before. The resulting metric is,

\begin{eqnarray}
\frac{ds^{2}}{\alpha^{'}} &=& C[W_{0}^{2}(-dt^{2}+\sum(dx^{i})^{2})+d\zeta^{2}+\cos^{2}\zeta d\psi^{2} \nonumber \\
&+& \frac{1}{2}\sin^{2}\zeta(d\phi_{1}^{2}+d\phi_{2}^{2})+W_{0}^{4}\sin^{2}\zeta dt(d\phi_{1}+d\phi_{2})] \ .
\end{eqnarray}
We shall look for string propagation in this background using
the following ansatz,
\begin{eqnarray}
t=t(\tau), \,\ x_{i}=x_{i}(\tau), \,\ \psi=\psi(\tau), \,\ \zeta=\zeta(\tau), \,\
 \phi_{1}=\phi_{1}(\tau)+m_{1}\sigma, \,\  \phi_{2}=\phi_{2}(\tau)+m_{2}\sigma .  \nonumber\\
\end{eqnarray}
Again we have to show that the above embedding is self-consistent
with the constraint equations as in the case before.
To check this, we start by solving the equations of motion using the ansatz above.
Solving the $t$ equation of motion we get,
\begin{equation}
     \ddot{t} = \frac{W_0^2}{2}\partial_{\tau}\{(\dot{\phi}_1 + \dot{\phi}_2)\sin^2\zeta\} \ .  \label{100}
\end{equation}
Solving for $\phi_1$ and $\phi_2$ respectively we get,
\begin{eqnarray}
    \dot{\phi}_1 &=& \frac{c_5}{\sin^2\zeta} - W_0^4\dot{t} \\ \nonumber \dot{\phi}_2 &=& \frac{c_6}{\sin^2\zeta} - W_0^4\dot{t} \ ,  \label{101}
\end{eqnarray}
where $c_5$ and $c_6$ are integration constants. Substituting the value of $\dot{\phi_1}$ and $\dot{\phi_2}$
from (\ref{101}) into (\ref{100}) we get,
\begin{eqnarray}
    \ddot{t} &=& 0 \>\>\> \Rightarrow \dot{t} = c_4 \ ,
\end{eqnarray}
where $c_4$ is the integration constant. Solving for $x_i$ and $\psi$ we get,
\begin{eqnarray}
    \dot{x_i} &=& c_i, \>\>\> \dot{\psi} = \frac{c_7}{\cos^2\zeta} \ .
\end{eqnarray}
So the equations for $W$ and $\theta$ generate the constraints
\begin{eqnarray}
    c_4^2 - \sum c_i^2 &=& 2W_0^2c_4(c_5+c_6), \nonumber \\ \frac{c_5^2-c_6^2}{\sin^4\zeta} &=& m_1^2-m_2^2 \ ,  \label{condition}
\end{eqnarray}
For the same reason as discussed in the previous section we must impose the constraint $c_5=c_6$ which implies $m_1^2=m_2^2$.
These conditions merely points out that $\dot{\phi_1} = \dot{\phi_2}$ (i.e. the corresponding angular momenta are equal)
and fix the values of $c_5$ and $c_6$ from the above equations.
Substituting these conditions into the $\zeta$  equation we get,
\begin{equation}
    \frac{d^2\zeta}{d\tau^2} = \sin\zeta \cos\zeta \Big[-\frac{c_7^2}{\cos^4\zeta} + \frac{c_5^2}{\sin^4\zeta} -m^2 \Big] \ .
\end{equation}
Integrating the above we arrive at
\begin{equation}
    \Big(\frac{d\zeta}{d\tau}\Big)^2 = -\frac{c_7^2}{\cos^2\zeta} - \frac{c_5^2}{\sin^2\zeta} - m^2\sin^2\zeta + c_8 \ , \label{EOM}
\end{equation}
where $c_8$ is an integration constant.
Now looking at the isometries of the background, we can evaluate the constants of motion from the action
as,
\begin{eqnarray}
E &=& \sqrt{\lambda}\mathcal{E}=\sqrt{\lambda}[W_{0}^{2}\dot{t}-
\frac{1}{2}W_{0}^{4}\sin^{2}\zeta(\dot{\phi}_{1}+\dot{\phi}_{2})]\ ,
\nonumber\\
P_{i} &=& \sqrt{\lambda}\mathcal{P}_{i}=\sqrt{\lambda}W_{0}^{2}\dot{x_{i}} \ , \nonumber\\ J_{\phi_{1}} &=& \sqrt{\lambda}\mathcal{J}_{\phi_{1}}=
\frac{\sqrt{\lambda}}{2}\sin^{2}\zeta[\dot{\phi_{1}}+W_{0}^{4}\dot{t}]\ , \nonumber\\
J_{\phi_{2}} &=& \sqrt{\lambda}\mathcal{J}_{\phi_{2}}=
\frac{\sqrt{\lambda}}{2}\sin^{2}\zeta[\dot{\phi_{2}}+W_{0}^{4}\dot{t}]\ , \nonumber\\
J_{\psi} &=& \sqrt{\lambda}\mathcal{J}_{\psi}=\sqrt{\lambda}\cos^{2}\zeta\,\dot{\psi} \ . \label{charges2}
\end{eqnarray}
Also, we can see that the second Virasoro constraint in this case implies
that
\begin{equation}
m_{1}\mathcal{J}_{\phi_{1}}+m_{2}\mathcal{J}_{\phi_{2}}=0 \ .
\end{equation}
Since in this calculation we will be interested in the subset of solutions
which have two equal spins i.e.
\begin{equation}
\mathcal{J}_{\phi_{1}}=\mathcal{J}_{\phi_{2}},\,\,\,\,\,\,\,\,\,\,\,\,\,\,\,\,\, \Rightarrow m_{1}=-m_{2}=m \ .
\end{equation}
We can see that this is in perfect agreement with the equation (\ref{condition}), thus
making our solutions completely consistent. Also,
here the first Virasoro constraint gives the evolution equation for
$\zeta$
\begin{equation}
\dot{\zeta}^{2}=W_{0}^{2}(\dot{t}^{2}-\dot{x_{i}}^{2})-\cos^{2}\zeta\,\dot{\psi}^{2}-\frac{1}{2}\sin^{2}\zeta
[\dot{\phi_{1}}^{2}+\dot{\phi_{2}}^{2}+2W_0^4(\dot{\phi}_{1}+\dot{\phi}_{2})\dot{t}+2m^{2}] \ , \label{EOM2}
\end{equation}
which can be shown to be exactly equivalent to (\ref{EOM}) with putting in the values
and the identification $c_8 = W_{0}^{2}(\dot{t}^{2}-\dot{x_{i}}^{2}) = W_0^2(c_4^2 - \sum c_i^2)$.
So, in this case we note that the constraint equations (\ref{condition}) are satisfied
completely without restricting our parameter space non-trivially as before.

Putting in the values from (\ref{charges2}) into (\ref{EOM2}), we get
\begin{equation}
\dot{\zeta}^{2}=\frac{\widetilde{E}^{2}}{W_{0}^{2}}-
\frac{\mathcal{J}_{\psi}^{2}}{\cos^{2}\zeta}-\frac{\mathcal{J}^{2}}{\sin^{2}\zeta}
-m^{2}\sin^{2}\zeta \ ,
\end{equation}
where $\widetilde{E}^{2}=\mathcal{E}^{2}-\sum\mathcal{P}_{i}^{2}+2W_{0}^{4}(\mathcal{J}_{\phi_{1}}+\mathcal{J}_{\phi_{2}})$
and $\mathcal{J}^{2}=2(\mathcal{J}_{\phi_{1}}^{2}+\mathcal{J}_{\phi_{2}}^{2}) \ ,$
so that $\mathcal{J}$ is a real quantity. Now the equation of motion
for $\zeta$ looks like the classical equation for a particle moving
in a potential. Notice that the potential here grows to infinity at
both $\zeta=0$ as well as $\zeta=\frac{\pi}{2}$. So the functional
form suggests a infinite potential well with a minimum in between
the extremas. The $\zeta$ coordinate must then oscillate in this
well between a maximum and minimum value. We define the Oscillation
number for the system as
\begin{equation}
\mathcal{N}=\frac{1}{2\pi}\oint\, d\zeta\,\dot{\zeta}=
\frac{1}{\pi}\int_{\zeta_{min}}^{\zeta_{max}}d\zeta
\sqrt{\frac{\widetilde{E}^{2}}{W_{0}^{2}}-
\frac{\mathcal{J}_{\psi}^{2}}{\cos^{2}\zeta}-
\frac{\mathcal{J}^{2}}{\sin^{2}\zeta}-m^{2}\sin^{2}\zeta} \ ,
\end{equation}
with $\mathcal{N}=\frac{N}{\sqrt{\lambda}}$ being an Adiabatic invariant,
which should have integer values in the usual quantum theory. Putting $\sin\zeta=x$ into the integral for oscillation number, we
get
\begin{equation}
\mathcal{N}=\frac{1}{\pi}\int_{\sqrt{R_{1}}}^{\sqrt{R_{2}}}\frac{dx}{1 - x^{2}}\sqrt{\frac{\widetilde{E}^{2}}{W_{0}^{2}}(1-x^{2})-\mathcal{J}_{\psi}^{2} - \frac{\mathcal{J}^{2}(1-x^{2})}{x^{2}}-m^{2}x^{2}(1-x^{2})} \ ,
\end{equation}
where $R_{1}$ and $R_{2}$ are two positive appropriate roots of
the polynomial
\begin{eqnarray}
 g(z) &=& m^{2}z^{3}+(-\frac{\widetilde{E}^{2}}{W_{0}^{2}}-m^{2})z^{2}
+
(\frac{\widetilde{E}^{2}}{W_{0}^{2}}+\mathcal{J}^{2}-\mathcal{J}_{\psi}^{2})z
-
\mathcal{J}^{2},\,\,\,\,\,\, z=x^{2}.
\end{eqnarray}

Naturally, we will be interested in the region of parameter space
where the roots to the above polynomial are real. Now taking the partial derivative of $\mathcal{N}$ w.r.t $m$ we
get
\begin{equation}
\frac{\partial\mathcal{N}}{\partial m}=-\frac{m}{\pi}\int_{\sqrt{R_{1}}}^{\sqrt{R_{2}}}dx\,
\frac{x^{3}}{\sqrt{\frac{\widetilde{E}^{2}}{W_{0}^{2}}(1-x^{2})-
\mathcal{J}_{\psi}^{2}-\frac{\mathcal{J}^{2}(1-x^{2})}{x^{2}}-m^{2}x^{2}(1-x^{2})}}
\end{equation}
Now to find the roots of the polynomial $g(z)$ we do an approximate
analysis. In the large $\tilde{E}$ but small $\mathcal{J}$ and $\mathcal{J}_{\psi}$
limit, we can find the three distinct roots as,
\begin{eqnarray}
\alpha_{1} &=& \frac{\widetilde{E}^{2}}{mW_{0}^{2}}+W_{0}^{2}\frac{\mathcal{J}_{\psi}^{2}-
\mathcal{J}^{2}}{\widetilde{E}^{2}}+\mathcal{O}[W_{0}^{4}\widetilde{E}^{-4}]\ , \nonumber\\
\alpha_{2} &=& \frac{W_{0}^{2}\mathcal{J}^{2}}{\widetilde{E}^{2}}+
\mathcal{O}[W_{0}^{4}\widetilde{E}^{-4}]\ , \nonumber\\
\alpha_{3} &=& 1-\frac{W_{0}^{2}\mathcal{J}_{\psi}^{2}}{\widetilde{E}^{2}}+
\mathcal{O}[W_{0}^{4}\widetilde{E}^{-4}]\ .
\end{eqnarray}
Clearly we can see, $0\leq x^{2}\leq1$ , so in the large $\widetilde{E}$
limit, we choose the appropriate upper and lower limit to the integral
accordingly. Putting $x^{2}=z$ we write the integral as
\begin{equation}
\frac{\partial\mathcal{N}}{\partial m}=-\frac{m}{2\pi}\int_{\alpha_{2}}^{\alpha_{3}}dz\,\frac{z}{\sqrt{m^{2}z^{3}+
(-\frac{\widetilde{E}^{2}}{W_{0}^{2}}-m^{2})z^{2}+
(\frac{\widetilde{E}^{2}}{W_{0}^{2}}+
\mathcal{J}^{2}-\mathcal{J}_{\psi}^{2})z-\mathcal{J}^{2}}} \ .
\end{equation}
Using standard integral tables we can transform this into a combination
of the usual Elliptic integrals of first and second kind as,
\begin{equation}
\frac{\partial\mathcal{N}}{\partial m}=-\frac{m}{\pi}\frac{1}{\sqrt{\alpha_{1}-
\alpha_{2}}}\left[\alpha_{1}\mathbf{K}\left(\frac{\alpha_{3}-
\alpha_{2}}{\alpha_{1}-\alpha_{2}}\right)-(\alpha_{1}-
\alpha_{2})\mathbf{E}\left(\frac{\alpha_{3}-\alpha_{2}}{\alpha_{1}-
\alpha_{2}}\right)\right] \ .
\end{equation}
We expand the equation again in the large $\widetilde{E}$ but small
$\mathcal{J}$ and $\mathcal{J}_{\psi}$ limit to get,
\begin{equation}
\frac{1}{W_{0}}\frac{\partial\mathcal{N}}{\partial m}=c_{1}m^{2}\widetilde{E}^{-1}+c_{2}m^{4}\widetilde{E}^{-3}\left[c_{3}+
\frac{\mathcal{J}^{2}-\mathcal{J}_{\psi}^{2}}{m^{2}}\right]+
\mathcal{O}[W_{0}^{5}\widetilde{E}^{-5}] \ ,
\end{equation}
 where the numercial constants are given by $c_{1}=c_{2}=-0.25$ and
$c_{3}=0.375$.
Integrating this equation we get a series for $\mathcal{N}$ ,

\begin{equation}
\mathcal{N=}\mathcal{N}_{0}+\frac{c_{1}}{3}m^{2}W_{0}\widetilde{E}^{-1}+
\frac{c_{2}}{5}m^{5}W_{0}\widetilde{E}^{-3}\left[c_{3}+
\frac{5}{3}\frac{\mathcal{J}^{2}-\mathcal{J}_{\psi}^{2}}{m^{2}}\right]+
\mathcal{O}[W_{0}^{5}\widetilde{E}^{-5}] \ .
\end{equation}
The integration constant $\mathcal{N}_{0}$ can be evaluated by considering
the integral for $m=0$, i.e.

\begin{equation}
\mathcal{N}_{0}=\frac{1}{\pi}\int_{\beta_{1}}^{\beta_{2}}\frac{dx}{1
-x^{2}}\sqrt{\frac{\widetilde{E}^{2}}{W_{0}^{2}}(1-x^{2})+
\mathcal{J}^{2}(1-\frac{1}{x^{2}})-\mathcal{J}_{\psi}^{2}} \ ,
\end{equation}
where the limits are given by
\begin{equation}
   \beta^{2}=\beta_{2,1}^{2}=\frac{-(\frac{\widetilde{E}^{2}}{W_{0}^{2}}
+\mathcal{J}^{2}-\mathcal{J}_{\psi}^{2})\pm\sqrt{(\frac{\widetilde{E}^{2}}{W_{0}^{2}}
+\mathcal{J}^{2}-\mathcal{J}_{\psi}^{2})^{2}-
4\frac{\widetilde{E}^{2}}{W_{0}^{2}}\mathcal{J}^{2}}}{-2
\frac{\widetilde{E}^{2}}{W_{0}^{2}}} \ .
\end{equation}
Now using $\frac{\widetilde{E}^{2}}{W_{0}^{2}}+\mathcal{J}^{2}-\mathcal{J}_{\psi}^{2}=
\frac{\widetilde{E}^{2}}{W_{0}^{2}}\beta^{2}+\frac{\mathcal{J}^{2}}{\beta^{2}}$
and changing the variable, we transform the integral to
\begin{equation}
\mathcal{N}_{0}=\frac{\beta_{1}}{\pi}\int_{1}^{\frac{\beta_{2}}{\beta_{1}}}
\frac{dx}{1-\beta_{1}^{2}x^{2}}\sqrt{\frac{\widetilde{E}^{2}}{W_{0}^{2}}
\beta_{1}^{2}(1-x^{2})+\frac{\mathcal{J}^{2}}{\beta_{1}^{2}}(1-\frac{1}{x^{2}})}
=\frac{1}{2}(\frac{\widetilde{E}}{W_{0}}-\mathcal{J}+\mathcal{J}_{\psi}) \ .
\end{equation}
We put back this value into and then, by reverting the series we get
\begin{eqnarray}
\frac{\widetilde{E}}{W_{0}} &=& 2\mathcal{N}+(\mathcal{J}-\mathcal{J}_{\psi})+a_{1}m^{3}\mathcal{N}^{-1}-
a_{2}m^{3}\mathcal{N}^{-2}(\mathcal{J}-\mathcal{J}_{\psi}) \nonumber \\ &+& a_{3}m^{6}\mathcal{N}^{-3}A(m,\mathcal{J},\mathcal{J}_{\psi})
- a_{4}m^{6}\mathcal{N}^{-4}(\mathcal{J}-
\mathcal{J}_{\psi})B(m,\mathcal{J},\mathcal{J}_{\psi})+
\mathcal{O}[\mathcal{N}^{-5}] \ , \nonumber \\
\end{eqnarray}
which reduces to the usual linear scaling relation of energy with
spins and oscillation number in the large $\mathcal{N}$ limit. Here $a_{1}\simeq0.08334$, $a_{2}\simeq0.04167$, $a_{3}\simeq0.00347$,
$a_{4}\simeq0.00521$ , and
\begin{eqnarray}
A(m,\mathcal{J},\mathcal{J}_{\psi}) &=& -1+\frac{d_{1}}{m}+\frac{d_{2}\mathcal{J}
(\mathcal{J}-\mathcal{J}_{\psi})}{m^{3}}\ , \nonumber\\
B(m,\mathcal{J},\mathcal{J}_{\psi})&=&
-1+\frac{d_{1}}{m}+\frac{d_{2}(2\mathcal{J}^{2}-
\mathcal{J}\mathcal{J}_{\psi}-\mathcal{J}_{\psi}^{2})}{3m^{3}}\ ,
\end{eqnarray}
with $d_{1}\simeq1.35$, $d_{2}\simeq12$. We can see that no higher
powers of $W_{0}$ appears in the series, so we can claim that our
approximation on $W_{0}$ does not bring any divergences in the
spectrum of $\widetilde{E}$. Also we recall that
$\widetilde{E}^{2}=\mathcal{E}^{2}-\sum\mathcal{P}_{i}^{2}+2W_{0}^{4}(\mathcal{J}_{\phi_{1}}+\mathcal{J}_{\phi_{2}})$,
and for the sake of completeness we compute the expansion for
$\widetilde{\mathcal{E}}=\sqrt{\mathcal{E}^{2}-\sum\mathcal{P}_{i}^{2}}$.
It is easy to find that the solution can be written as,
\begin{equation}
\widetilde{\mathcal{E}}=2\mathcal{N}W_{0}+(\sqrt{2(\mathcal{J}_{\phi_{1}}^{2}+
\mathcal{J}_{\phi_{2}}^{2})}-\mathcal{J}_{\psi})W_{0}+
\sum_{n=1}^{\mathcal{1}}(\frac{1}{\mathcal{N}})^{n}
\mathcal{G}_{(n)}(m,\mathcal{J}_{\phi_{1}},\mathcal{J}_{\phi_{2},}\mathcal{J}_{\psi}) \ ,
\end{equation}
where
\begin{eqnarray}
\mathcal{G}_{(k)}&=&[f_{1}(m,\mathcal{J},\mathcal{J}_{\psi})W_{0}+
(\mathcal{J}_{\phi_{1}}+\mathcal{J}_{\phi_{2}})f_{2}(m,\mathcal{J},
\mathcal{J}_{\psi})W_{0}^{3} \nonumber \\
&+&(\mathcal{J}_{\phi_{1}}+\mathcal{J}_{\phi_{2}})^{2}f_{3}(m,
\mathcal{J},\mathcal{J}_{\psi})W_{0}^{5}
+(\mathcal{J}_{\phi_{1}}+\mathcal{J}_{\phi_{2}})^{3}f_{4}(m,
\mathcal{J},\mathcal{J}_{\psi})W_{0}^{7}+...] \ . \nonumber \\
\end{eqnarray}
Here the functions $f_{k}(m,\mathcal{J},\mathcal{J}_{\psi})$ are of rather complicated form and we do
not present them here explicitly. But again, it seems clear that even
without terms higher than $\mathcal{O}(W_{0}^{4})$ the series does
not show any divergences, hinting at a well behaved energy spectrum.
\section{Conclusion} In this paper we have studied few examples of
semiclassical strings in the near horizon geometry of PFT. We have
found the most general solutions of the equations of motion of the
probe fundamental strings in this background and found out
dispersion relations among various conserved quantities using some
regularization technique. However, while studying semiclassical
strings in PFT background we have used some simplification and
kept terms upto $\mathcal{O}(V^4)$, where $V$ is the radial
coordinate. This approximation is justified by following
\cite{Ganor:2007qh}, which would correspond to the leading order
deformation to $N=4$ SYM. Also putting $W$ and $\theta$ to be constants
has made us run into non-trivial constraints on the parameter space.
We can try to study string propagation
in the background with full generality. It will also be highly
challenging to study the boundary theory operators corresponding
to these states as the dual gauge theory is almost unknown beyond
the leading order. Hence, the semiclassical analysis of the string
states might give us hints about the possible nature of dual gauge
theory operators next to leading order. Furthermore it will be
interesting to study the Wilson loops in this background to have a
better understanding of this. We hope to come back to some of
these issues in future.

\section*{Acknowledgement} A. B. would like to thank P.M. Pradhan for
some useful discussions.

\end{document}